\documentclass[a4paper,12pt]{article}
\usepackage[dvips]{graphicx}
\usepackage{amsmath}
\def\tmg{\tau \to\mu\gamma}
\def\meg{\mu \to e\gamma}
\def\Brt{Br(\tau \to\mu\gamma)}
\def\Brm{Br(\mu \to e\gamma)}
\def\tanB{\tan\beta }
\def\10{\bf 10}
\def\5b{\bf {\bar{5}}}

\def\xn{\tilde{\chi}^0}

\begin{document}

\begin{flushright}
hep-ph/0612370\\
DPNU-06-10\\
December 30, 2006
\end{flushright}

\vspace{4ex}

\begin{center}
{\large \bf
Lepton Flavor Violation in SUSY GUT Model \\ with Non-universal Sfermion Masses
}

\vspace{6ex}

\renewcommand{\thefootnote}{\alph{footnote}}
S.-G. Kim\footnote{e-mail: sunggi@eken.phys.nagoya-u.ac.jp},
N. Maekawa\footnote{e-mail: maekawa@eken.phys.nagoya-u.ac.jp},
A. Matsuzaki\footnote{e-mail: akihiro@eken.phys.nagoya-u.ac.jp}, 
\\K. Sakurai\footnote{e-mail: sakurai@eken.phys.nagoya-u.ac.jp},
and 
T. Yoshikawa\footnote{e-mail: tadashi@eken.phys.nagoya-u.ac.jp}

\vspace{4ex}
{\it Department of Physics, Nagoya University, Nagoya 464-8602, Japan}\\

\end{center}

\renewcommand{\thefootnote}{\arabic{footnote}}
\setcounter{footnote}{0}
\vspace{6ex}


\begin{abstract}

We analyze lepton flavor violating $\tmg$ and $\meg$ processes in SUSY GUT model in which sfermions have special mass spectrum.
It is assumed that only third generation sfermions which are contained in ${\bf 10}{\rm (Q, U^c, E^c)}$ of SU(5) can have a different mass from the others.
This mass spectrum is led from ${\rm E_6}$ GUT model with horizontal symmetries.
It is shown that branching ratios of $\tmg$ and $\meg$ depend strongly on a right-handed stau mass.
The weak scale stability requires the light stau, so large decay rates can be expected in this scenario.
When the stau mass is around 150 GeV and $\tanB\sim 10$, the branching ratios can be larger than $\Brt \simeq 10^{-8}$ and 
$\Brm \simeq 5\times 10^{-12}$, which are within reach of future experiments.
In addition, this model has an interesting feature that the final state charged lepton tends to have the 
right-handed chirality.

\end{abstract}

\vspace{1cm}


\section{Introduction}

The measurement of the Lepton Flavor Violation (LFV) in charged lepton sector is one of the most important measurements to search for New Physics beyond the Standard Model (SM).
Since the neutrino oscillations have been observed, the charged LFV processes can be expected to be non-vanishing.
However, the SM with the Majorana neutrino masses leads too small LFV decay rates to be observed due to smallness 
of the neutrino masses\cite{LFVSM}.
Thus, detections of the LFV processes means discovery of New Physics. 

In fact the supersymmetric extension of the SM leads generically 
large LFV decay rates, because of picking up the off-diagonal entries of 
slepton mass matrices (in substitution for neutrinos) in intermediate state.  
However, it is well known that Flavor Changing Neutral Current (FCNC) and CP violating processes are highly suppressed experimentally.
They impose severe constraints on the minimal supersymmetric (SUSY)
standard model (MSSM) parameter space\cite{Masiero}.
The simplest way to satisfy these FCNC constraints is to adopt the universal sfermion soft mass matrices at some scale.
       
In many analysis that have been done so far, the universal soft masses at the gravitational scale or at the grand unified scale have been assumed and the LFV decay rates are calculated by the non-universality which is induced by the radiative correction\cite{LFVn,LFV10}.
For example, in the MSSM with the Majorana neutrino masses,
the radiative correction induces the non-universal sfermion masses\cite{LFVn}.
Unfortunately, it is difficult to predict the LFV decay rates quantitatively in
this scenario unless the Dirac neutrino Yukawa couplings are fixed,
because such LFV decay rates strongly depend on the Dirac neutrino Yukawa
 couplings, which looks impossible to be measured directly.
In  ${\rm SO(10)}$ grand unified theory (GUT), the Dirac neutrino Yukawa 
couplings are related with the up-type quark Yukawa couplings. Then
we can predict the LFV decay rates\cite{LFV10}, but they are too large to be consistent
with the experimental bound if the diagonalizing matrix of the fermions
mass matrices for doublet lepton has the Maki-Nakagawa-Sakata\cite{MNS}(MNS)-{\it like} large mixings.  
In this direction, some special structure which suppresses the $\mu\rightarrow e\gamma$ process is required.
 
The assumption of universal soft masses is not necessarily required. 
For example, since the FCNC constraints for the first two generation fields are much more severe than for the third generation fields, the universal soft masses sometimes have been imposed only for the first two generation sfermion,
 which are realized if we impose non-Abelian  
horizontal symmetry, for example, ${\rm U(2)}$, under which the first two generation fields are
doublets and the third generation fields are singlets\cite{horizontal}.
However, if the diagonalizing matrix of the fermions have the MNS-{\it like} large mixings, 
such sfermion mass spectrum leads to too large FCNC to be consistent with the experimental bound.
Therefore, such non-universality must be introduced only for the sfermion whose fermionic superpartner
has the diagonalizing matrix with the Cabibbo-Kobayashi-Maskawa\cite{CKM}(CKM)-{\it like} small mixings. 

In the SU(5) SUSY-GUT, it is a reasonable assumption that 
the diagonalizing matrices of ${\bf 10}$ 
representation fields have the CKM-{\it like} small mixings, while those
of ${\bf\bar 5}$ fields have the MNS-{\it like} large mixings,
because ${\bf 10}$ representation 
includes the doublet quarks, and ${\bf\bar 5}$ includes the doublet
leptons.
Under this assumption, only the third generation sfermions of the ${\bf 10}$ representation can have different masses from the others without contradicting with experimental constraints from various FCNC processes.
In this paper, we study the LFV under the non-universal sfermion masses above.

Before examining the LFV, we summarize several characteristic features of the non-universal sfermion masses.
First of all, 
the rotation matrix for $\10$ which makes fermions to mass eigenstate is expected to have the CKM-{\it like} small mixings, since $\10$ involves quark doublet.
Therefore, large off-diagonal entries for $\10$ representation sfermion mass matrix do not arise after this rotation, even if the initial soft masses are not degenerate.
Moreover, FCNC constraints among 1-3 or 2-3 generations are not so severe than that of 1-2 generations.           
Therefore, we can expect that non-degeneracy for third generation does not conflict the FCNC constraints.
And it is important that we can expect larger FCNC for the third generation fields than in the usual 
universal sfermion mass case.
Second, the naturalness of the Higgs mass in MSSM requires that the gaugino masses, the higgsino mass,
 and the stop masses 
must be around the weak scale. In the above non-universality case, we can take the larger masses for
the other sfermion masses than the third generation sfermions of the ${\bf 10}$ representation without
conflicting the naturalness arguments,
because the both stops (left and right handed stops) are included in the third generation
${\bf 10}$ representation of ${\rm SU(5)}$. In principle, we can take such mass larger than 1 TeV, which
can weaken the various constraints from FCNC, electric dipole moments (EDM) and the muon $g-2$, etc. related with the first two generation fields.
Again, the FCNC related with the third generation fields can become relatively large, which may be detectable in future experiments.
Finally, such non-universality of the sfermion masses is naturally realized in the ${\rm E}_6$ 
SUSY-GUT model\cite{E6, E6r} with ${\rm SU(2)}$ or ${\rm SU(3)}$ horizontal symmetry\cite{E6Hori}.

In this paper, we will analyze the lepton flavor violating $\tmg$ and $\meg$ processes under this non-universal sfermion masses. The point is that the naturalness requires that the ${\bf 10}_3$ sfermion masses are ${\cal O}(\Lambda_W)$, i.e., 
the right-handed stau mass becomes ${\cal O}(\Lambda_W)$. 
We show that if mass of right-handed stau 
is near the weak scale, each LFV decay rate 
becomes sufficiently large to be found by near future experiments.
Actually, the LFV decay rates are very sensitive to the right-handed stau mass.
Thus, if $\tmg$ or $\meg$ processes are discovered and the branching ratios are measured, this model may predict a mass of right-handed stau.
We will see that this model has interesting feature, that is, this model indicates that
the  final state charged lepton tends to be right-handed, since main source of flavor violation comes from the Feynman diagram which includes the right-handed charged slepton.

This paper is organized as follows.              
In the next section, we will specify the model, and discuss how lepton flavor violating processes take place in this model.
The results of our numerical study are given in Section 3.
In Section 4, we will give a summary and discussion of our results.

\section{Lepton Flavor Violation in this model}

In this section, we discuss how lepton flavor violating decays take place in a SUSY-GUT model 
with the non-universal sfermion masses mentioned in the previous section.
First, we assume that the SM gauge group $\rm SU(3)_C \times SU(2)_L \times U(1)_Y$ is unified into
 ${\rm SU(5)}$. (Note that the following arguments can be applied to unified models with larger unified
 gauge groups, for instance, ${\rm SO(10)}$ or ${\rm E_6}$.) 
Then, around the GUT scale, the Lagrangian for the SUSY breaking terms is
given as
\begin{eqnarray}
-{\cal L}_{soft~breaking} &=& -{\cal L}_{soft~mass} 
+ \bigl( A^u_{ij} \tilde{\bf 10}_i \tilde{\bf 10}_j {\bf 5}_H
+ A^{d,e}_{ij} \tilde{\bf 10}_i \tilde{\bf \overline{5}}_j {\bf \overline{5}}_H
\nonumber \\
&~&~~~~~~~~~~~~~~~~~+ A^{\nu}_{ij} \tilde{\bf 1}_i \tilde{\bf \overline{5}}_j {\bf 5}_H
+ B {\bf \overline{5}}_H {\bf 5}_H + \frac{1}{2} M_{1/2} \lambda^a \lambda^a + {\rm h.c.} \bigr)\,,\\
\label{A term}
-{\cal L}_{soft~mass} &=& (m^2_{\bf 10})_{ij} \tilde{\bf 10}^{\dagger}_i \tilde{\bf 10}_j 
+ (m_{\bf\bar 5}^2)_{ij} \tilde{\bf \overline{5}}_i^{\dagger} \tilde{\bf \overline{5}}_j
+ m^2_{\tilde{\nu}_Rij} \tilde{\bf 1}_i \tilde{\bf 1}_j
+ m^2_{H_d} {\bf \overline{5}}_H^{\dagger} {\bf \overline{5}}_H
+ m^2_{H_u} {\bf 5}_H^{\dagger} {\bf 5}_H
\label{soft mass L}\,,
\end{eqnarray}
where the tilded fields are regarded as sfermions, $\lambda^a$ is SU(5) gauginos, 
${\bf 10}_i = (Q_i, \bar{U}_i, \bar{E}_i)$,
${\bf \bar{5}}_i = (\bar{D}_i, L_i)$, ${\bf 1}_{i} = {\bar{N}}_i$ ($i=1,2,3$), 
$M_{1/2}$ is SU(5) gaugino mass, and $-{\cal L}_{soft~mass}$ gives sfermions and Higgs soft mass terms. ${\bf\bar 5}_H$
and ${\bf 5}_H$ include the down-type Higgs $H_d$ and the up-type Higgs $H_u$, respectively.

As mentioned in Introduction, we consider following forms for the sfermion mass matrices. 
\begin{equation}
m_{\bf 10}^2 = \begin{pmatrix}
m_0^2&&\\
&m_0^2&\\
&&m^2_{30}
\end{pmatrix},
~~~~~~~~~~~~
m_{\bf\bar 5}^2 = \begin{pmatrix}
m^2_0 &&\\
&m^2_0 &\\
&&m^2_0
\end{pmatrix}.
\label{special mass}
\end{equation}
The reasons are as follows:
\begin{itemize}
\item{\bf Severe Constraints on the first two Generation:}
\\
There are severe constraints on the off-diagonal entries of low energy sfermion mass matrices from
the various FCNC processes in the basis of the superfields in which all quark and lepton mass matrices are diagonal (
super-CKM basis).
This is famous as the SUSY-Flavor problem. 
In particular, the (1,2) component suffers from strong constraints by CP violating parameters of $K^0$-$\overline{K}{}^0$ mixing, lepton flavor violating $\meg$, and so on.
Thus, we cannot release degeneracy between 1-2 generation, otherwise a large (1,2) element arises 
in super-CKM basis and it conflicts with experiments.

\item{\bf Large Neutrino Mixing and Constraints on 2-3 Generation:}   
\\
It is known that there is a maximal mixing between 2-3 generation in lepton sector\cite{Atm}.
This large mixing sometimes is related with the mixing of $\5b$ since $\5b$ contains left-handed lepton doublet. 
Thus, if we release degeneracy between 2-3 generations in $\5b$, a large 2-3 component of $\bf\tilde{\5b}$ is generated in super-CKM basis.
It conflicts with constraints from $B^0$-$\overline{B}{}^0$ mixing and $\tmg$.
Therefore, we should assume degeneracy of 2-3 generation for $\bf\tilde{\5b}$ soft masses.
While, we can take a soft mass of the third generation of $\bf\tilde{10}$ different from that of the other generation, since a mixing of $\10$ is expected to be as small as CKM mixing.

\item{\bf Naturalness of the MSSM:}
\\
In the MSSM, there is no terrible fine tuning problem due to supersymmetric cancellation for quadratic divergence for radiative correction to Higgs mass square.
Instead of the quadratic divergence, the radiative corrections are proportional to the SUSY breaking
mass square, and therefore SUSY breaking mass scale must be around the weak scale $\Lambda_W$ to avoid some tuning
if the couplings
are $O(1)$. Among the various sfermion masses, only the stop masses must be the weak scale, 
while the other sfermion masses can be taken much larger than the 
weak scale because of the smallness of the Yukawa couplings.
One of the advantages of the mass spectrum we assumed is that we can take all the sfermion masses much
larger than the weak scale except the masses of sfermions which are contained in ${\bf 10}_3$.
In such parameter region, several constraints in SUSY models, for example $g-2$, EDM, FCNC, etc, are 
relaxed without loss of naturalness.
Especially, A-term contribution to EDM or FCNC for the first two generations can be decoupled in this region.
In the literature, it is argued that the stop masses are required to be larger than the weak scale 
to realize the Higgs mass larger than the lower bound $m_h>114.4$ GeV obtained at LEP II.
Such large stop masses lead to some tuning in the Higgs sector. 
This is called ``the little hierarchy problem"\cite{little}. 
However, as discussed in \cite{KMMSSY}, the stop masses around 300-400 GeV can be consistent with the LEP 
II experiments, if the lightest CP-even Higgs has a small $ZZh$ coupling.
Therefore, some of the parameter region considered in this paper may lead to such small stop masses,
but it is consistent with the LEP II results of the Higgs search experiment.

\item{\bf ${\rm E_6}$ GUT with U(2) horizontal Symmetry:}
\\
Introducing a non-Abelian horizontal symmetry is one of various interesting possibilities to realize
the flavor blind nature of the sfermion masses\cite{horizontal,SU3}. 
Most of the previous attempts adopt U(2) or SU(2)$\times$ U(1) symmetry as the horizontal symmetry\cite{horizontal}, which
realize the universality of the first two generation sfermion masses and the large top Yukawa coupling,
if we assign doublets for the first two generation fields and singlets for the third generation
fields and the Higgs fields. Unfortunately, the universality of the first two generation fields is not
sufficient to suppress the FCNC constraints especially when the diagonalizing matrix of fermion mass 
matrices have the MNS-{\it like} large mixings as we pointed out in the previous section.
${\rm E_6}$ GUT $\times$ U(2) Horizontal Symmetry model\cite{E6Hori} succeeds to give not only the observing large
hierarchy of fermion masses but also the sufficient
 universality of the sfermion masses to avoid FCNC or EDM constraints in a simple way even when the diagonalizing matrix has the MNS-{\it like} large mixings.
It is essential that this model can naturally realize the sfermion mass spectrum which we discussed in 
this paper because of an ${\rm E_6}$ nature. 

\end{itemize}  
The superpotential is given at the GUT scale as follows. 
\begin{equation}
W = Y^u_{ij} {\bf 10}_i {\bf 10}_j {\bf 5}_H 
+ Y^{d,e}_{ij} {\bf 10}_i \overline{\bf 5}_j \overline{\bf 5}_H 
+ Y^{\nu}_{ij} {\bf 1}_i \overline{\bf 5}_j {\bf 5}_H
+ \frac{1}{2} M_{ij} {\bf 1}_i {\bf 1}_j + \mu \overline{\bf 5}_H {\bf 5}_H\,.
\label{SU5superpotential} 
\end{equation}
Since {\bf 10} contains left-handed quark doublet,
the diagonalizing matrices of {\bf 10} fields are naively expected as
\begin{equation}
{\bf 10}_i \rightarrow (V_{10})_{ij} {\bf 10}_j\,,~~~~~~~
V_{10} \sim \begin{pmatrix}
1&\lambda &\lambda^3\\
\lambda &1&\lambda^2\\
\lambda^3 &\lambda^2 &1
\end{pmatrix}\,,
\label{V10}
\end{equation}
where $\lambda$ is Cabibbo angle $(\lambda = 0.22)$ and we omit the ${\cal O}(1)$ coefficients.
This rotation generates off-diagonal entries of the sfermion masses of ${\bf 10}$ fields as
\begin{equation*}
V^{\dagger}_{10} \begin{pmatrix}
m^2_0&&\\
&m^2_0&\\
&&m^2_{30} \end{pmatrix}
V_{10}
~\sim~
\begin{pmatrix}
m^2_0&&\\
&m^2_0&\\
&&m^2_{30} 
\end{pmatrix}
+ m^2_{mix0},
\end{equation*}
\begin{equation}
m^2_{mix0} \sim (m^2_{30}-m^2_0) \begin{pmatrix}
\lambda^6&\lambda^5&\lambda^3\\
\lambda^5&\lambda^4&\lambda^2\\
\lambda^3&\lambda^2&0
\end{pmatrix},
\label{mixsource} 
\end{equation}
at the GUT scale.

In the following, 
we examine the LFV processes with the above sfermion masses.
We use the mass insertion method to estimate the amplitudes for a while.
As the slepton mass matrices at the SUSY breaking scale, 
we adopt the slepton mass matrices given by replacing the $m_0$, $m_{30}$, and $m_{mix0}$ with $m$, $m_3$, and
$m_{mix}$, respectively, which are the slepton mass
parameters at the SUSY breaking scale. As mensioned in Introduction, we neglect the radiatively induced off-diagonal
slepton mass matrices. 
The weak scale stability requires that 
$m_{3}$, $M_{1/2}$, $\mu$, and $A$ must be around the weak
scale, but the other parameters, for example, $m$, can be much larger than the weak scale. 
In this paper, we often take $m_{3}$, $M_{1/2}$, $\mu$, $A \ll  m$.
For the left-right mixing SUSY breaking parameters, $A$, we just neglect the contributions in this paper.
One reason is that this contributions are decoupled in the limit 
$m_{3}$, $M_{1/2}$, $\mu$, $A \ll m$, and the other reason is that we would like to examine the model
independent parts while the left-right mixing SUSY breaking parameters are strongly dependent on the 
explicit models.
On the other hand, the diagrams (a) and (b) are not decoupled because $m_{mix}^2$ becomes large in the limit.
Therefore, we consider only (a) and (b) for a while. 
\begin{figure}[h!]
\begin{tabular}{cc}
\begin{minipage}{0.5\hsize}
\begin{center}
\includegraphics[width=5.7cm,clip]{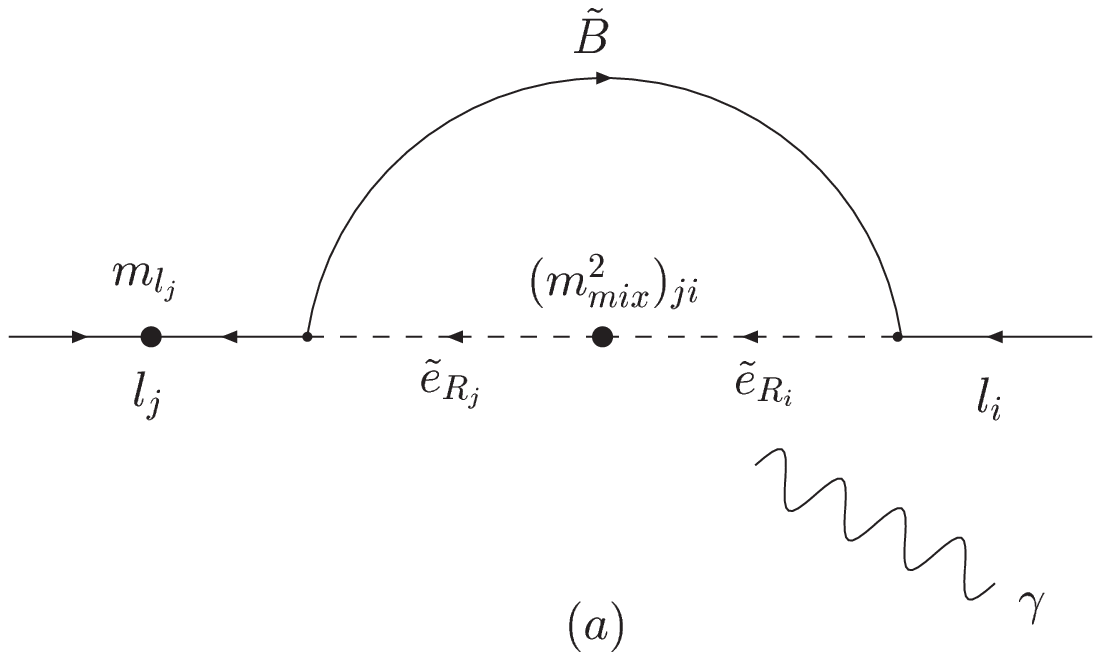}
\end{center}
\end{minipage}
\begin{minipage}{0.5\hsize}
\begin{center}
\includegraphics[width=5.7cm,clip]{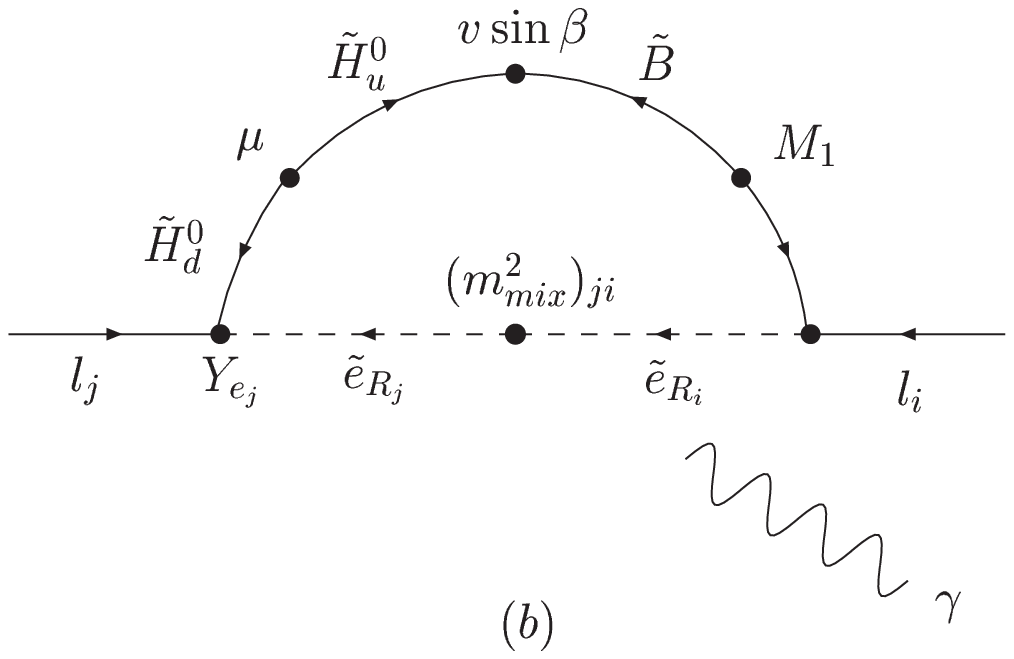}
\end{center}
\end{minipage}
\end{tabular}
\end{figure}
When $\tanB$ is large, the diagram (b) which is proportional to $Y_{e_j}=\sqrt{2}m_{e_j}/v\cos\beta$
is significant, where $v=246$ GeV.
The ratio of (b) and (a) is roughly 
\begin{equation}
\frac{{\rm Amp}(b)}{{\rm Amp}(a)} \sim \tanB\,.
\end{equation}\\
Let us roughly estimate the branching ratio of $\meg$ from the diagram (b) which is expected to give the largest contribution.
Comparing the diagram for $\meg$ with that for $\mu \to e\nu \bar{\nu}$ ($Br(\mu \to e\nu \bar{\nu}) \simeq 100 \%$),
we can roughly estimate the $\meg$ branching ratio as
\begin{eqnarray}
\Brm &\simeq& Br(\mu \to e\nu \bar{\nu}) \frac{\Gamma(\meg)}{\Gamma(\mu \to e\nu \bar{\nu})} \sim 
\frac{1}{16\pi^2} \frac{1}{G^2_F} \Bigl| e(g')^2 \frac{(\tan \beta) (m^2_{mix})_{21}}{m^4_S}  \Bigr|^2
\nonumber \\ 
&\sim& 
\frac{4\pi \alpha (\alpha')^2}{G^2_F} \frac{(\tan \beta)^2 \lambda^{10}}{m^4_S}\,,
\label{branch(m)}
\end{eqnarray}\\
where $G_F$ is Fermi constant and $m_S$ is typical superparticle mass.
For example, if we take $m_S = 400$GeV,~$\tanB = 5$, (\ref{branch(m)}) is found as
\begin{equation}
\Brm \sim 10^{-11} ~~(\,\times {\cal O}(1)^4~)~.
\end{equation}
We find that it is as large as the experimental upper bound ($\Brm<1.2 \times 10^{-11}$\cite{MEG}).
Since there is no lepton mass dependence in equation (\ref{branch(m)}), it can be applied for the $\tmg$ branching ratio.
Substituting $(m^2_{mix})_{32}$ and $Br(\tau \to e \nu \bar{\nu})$  for $(m^2_{mix})_{21}$ and $Br(\mu \to e \nu \bar{\nu})$, respectively, the $\tmg$ branching ratio is found roughly as
\begin{eqnarray}
\Brt &\sim& \Bigl( \frac{Br(\tau \to e \nu \bar{\nu})}{Br(\mu \to e \nu \bar{\nu})} \frac{\lambda^4}{\lambda^{10}} \Bigr) \times \Brm \nonumber \\
&\sim& 10^3 \times \Brm ~~~(\,\times {\cal O}(1)^4~)\,.
\end{eqnarray}
Then, for the same $m_S$ and $\tanB$ as in the above estimation of $Br(\meg)$,
\begin{equation}
\Brt \sim 10^{-8}~~~(\,\times {\cal O}(1)^4~)\,.
\end{equation}
The branching ratio of $\tmg$ is also as large as reach of experimental bound ($\Brt<4.5 \times 10^{-8}$\cite{TMG}).
The relation between $\Brt$ and $Br(\tau\rightarrow e\gamma$),
\begin{equation}
Br(\tau\rightarrow e\gamma)\sim\lambda^2\Brt~~~(\,\times {\cal O}(1)^4~)\,,
\end{equation}
 is also an interesting prediction of this scenario.
Unfortunately $\tau\rightarrow e\gamma$ process is more difficult to be observed because of the small branching ratio
because the experimental bound, $Br(\tau\rightarrow e\gamma)<1.1\times 10^{-7}$\cite{TMG}, is similar to that of 
$\tau\rightarrow \mu\gamma$ process.
Since the above estimations are very rough, more accurate calculations are necessary.
Such accurate calculations and the results are given in the next section. 

Before going to the numerical calculations, 
we mention about the other source of LFV. 
It has been discussed that the loop corrections induce the off-diagonal elements
of left-handed slepton soft masses\cite{LFVn} as
\begin{equation}
(m^2_{\tilde{l}})_{ij} \simeq -\frac{1}{8\pi^2}(2\bar{m}_0^2 + m^2_{H_u})(Y^{\nu\dagger}Y^{\nu})_{ij}
\log \frac{M_{mediated}}{M}\,,
\label{m2l}
\end{equation}
where $M$ and $M_{mediated}$ are the right-handed neutrino mass scale and the mediation scale of SUSY breaking, respectively.
If
\begin{equation}
(Y^{\nu\dagger}Y^\nu)_{23}\geq\lambda^2~~~~~~~~{\rm or}~~~~~~~~~(Y^{\nu\dagger}Y^\nu)_{12}\geq\lambda^5,
\label{condition1}
\end{equation}
then this loop correction may become sizable. For example, in SO(10) GUT model with minimal Higgs sector above equality is just hold because of SO(10) GUT relation $Y^{\nu} = Y^{u}$, though not in ${\rm E_6}$ models due to the smaller Dirac neutrino Yukawa couplings.
Since these contributions are decoupled again in the limit
$m_{3}$, $M_{1/2}$, $\mu$, $A \ll m$, we neglect this effect in this paper.

In case where $A^{d,e}$ have non-vanishing off-diagonal elements in the $Y^{d,e}$ diagonal basis, it leads large LFV.
The LFV occurs through the diagram (c).     
\begin{figure}[h!]
\begin{tabular}{cc}
\begin{minipage}{0.5\hsize}
\begin{center}
\includegraphics[width=5.7cm,clip]{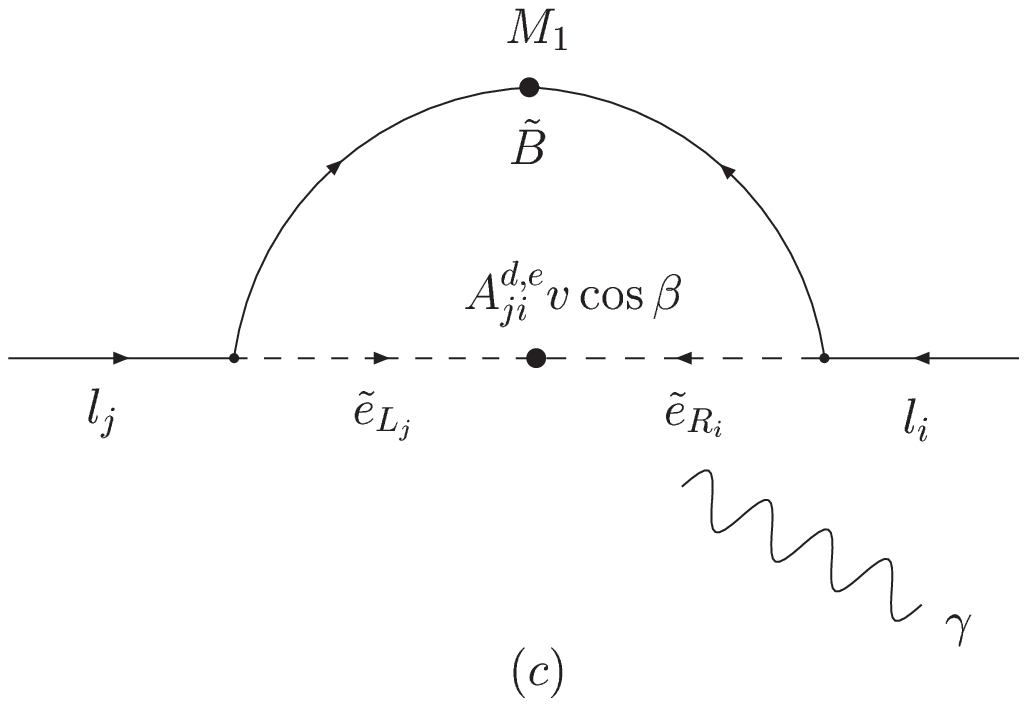}
\end{center}
\end{minipage}
\hspace{-1.7cm}
\begin{minipage}{0.5\hsize}
\begin{center}
\includegraphics[width=5.7cm,clip]{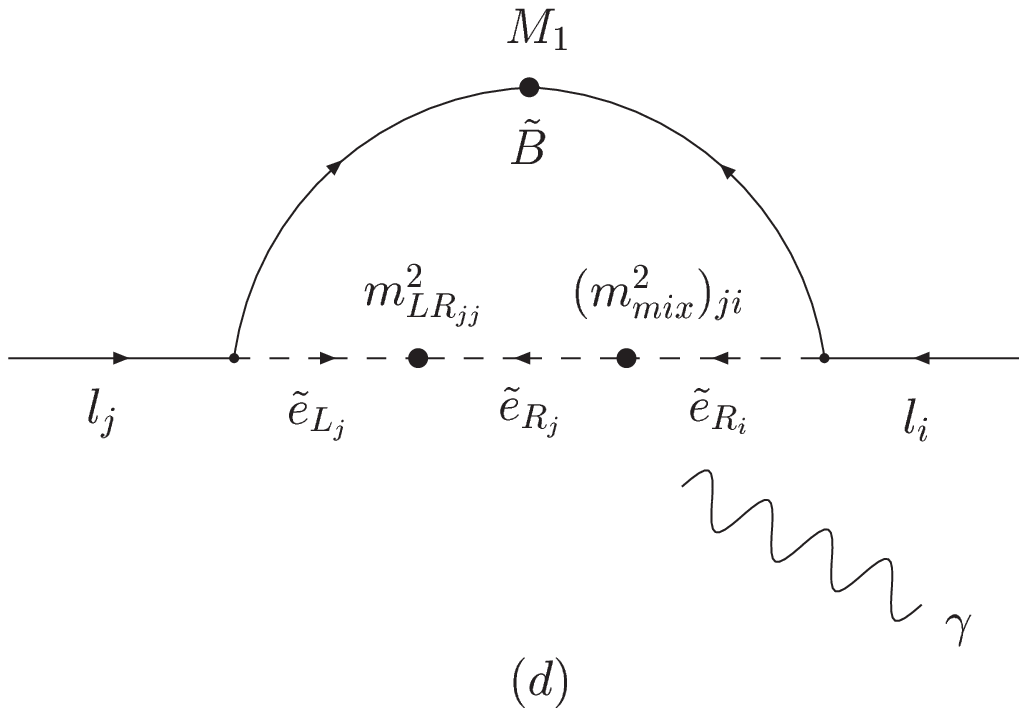}
\end{center}
\end{minipage}
\end{tabular}
\end{figure}
The ratio of amplitudes of diagram (c) and that of (b) is roughly found as 
\begin{equation}
\frac{{\rm Amp}(c)}{{\rm Amp}(b)} \sim \frac{M_1 A^{d,e}_{ij} v\cos\beta}{m_{l_j}\tanB (m^2_{mix})_{ji}}~.
\end{equation}
If $A^{d,e}_{ji} > \Hat{Y}^e_j \tanB (m^2_{mix})_{ji}/M_1$ is hold, a contribution from diagram(c) is more significant than (b). As we noted, this contribution is dependent on the explicit models and decoupled in the limit 
$m_{3}$, $M_{1/2}$, $\mu$, $A \ll m$, and therefore, we just neglect this effect in this paper.
When the A-terms are proportional to relevant Yukawa matrices ($A^{d,e} = a_0 Y^{d,e}$) or even when the A-term
is vanishing, the double-mass insertion diagram (d) via $\mu$ parameter can contribute the LFV processes. 
This contribution is included in our calculation in the next section, but it is small.
This may be because it is decoupled in the above limit and it is vanishing when $m_{3}=m$.

\section{The Branching Ratios}

Let us discuss the branching ratio of $\tau \to \mu \gamma$ and $\mu \to e \gamma$.
The amplitude of $l_j \to l_i \gamma$ is written as
\begin{equation}
T = \epsilon^{\alpha}(q) \bar{u}_i(p-q) i \sigma_{\alpha \beta} q^{\beta} (A_L^{(ji)}P_L + A_R^{(ji)}P_R)u_j(p)~,
\label{operator}
\end{equation}
where $p$ and $q$ represent the momenta of $l_j$ and photon, respectively.
Ignoring a mass of $l_i$ the decay rate is given by
\begin{equation}
\Gamma(l_j \to l_i \gamma) = \frac{m^3_{l_j}}{16\pi}(|A_L^{(ji)}|^2+|A_R^{(ji)}|^2)~.
\end{equation}
Now we can see an interesting feature in this model. 
This model says final state lepton tends to have the right-handed chirality.
From the operator form (\ref{operator}), we find that chirality flip is necessary with picking up the Yukawa couplings. 
In this model, only flavor violating coupling in the charged lepton sector is $m^2_{mix}$.
So, the intermediate state must have the right-handed chirality.
If the final state lepton has the left-handed chirality, we must pick up the Yukawa couplings at the final vertex ($Y_i$). 
On the other hand, if the final state has the right-handed chirality (then the initial state has the left-handed chirality)
 we must pick up that at the initial vertex ($Y_j$).
It means $A_L$ is as large as $A_R$ times $m_j / m_i$, so we find the final state lepton tends to have the right-handed
chirality.
We can test this prediction experimentally by measuring the angular distribution of final lepton.

In the basis where the charged slepton and the neutralino are mass eigenstates, 
the amplitude of $l_j \to l_i \gamma$ is given by
\begin{eqnarray}
A_L^{(ji)} \,&=&\, \frac{eg^2}{16\pi^2}\sum_{k,a}\frac{1}{m_{\tilde{e}_{k}}^2} \Bigl[
\sqrt{2} m_{l_j}(\sqrt{2} G^{akj}_{0eR}-H_{0eL}^{akj}) G^{*aki}_{0eR} F_2 \Bigl( \frac{M^2_{\tilde{\chi}^0_a}}{m^2_{\tilde{e}_{k}}} \Bigr) \nonumber\\
&&-\sqrt{2} M_{\tilde{\chi}^0_a} (\sqrt{2}G_{0eL}^{akj}+H^{akj}_{0eR}) G^{*aki}_{0eR} F_4
\Bigl( \frac{M^2_{\tilde{\chi}^0_a}}{m^2_{\tilde{e}_{k}}} \Bigr) \Bigr]\,, 
\label{MassEigenAmp}
\end{eqnarray}
where $m_{\tilde{e}_{k}}$ ($k=1,2,\cdots,6$) and $M_{\tilde{\chi}^0_a}$ $(a=1,\cdots,4)$ denote
 the masses of the charged slepton 
and the neutralino masses, respectively. 
$G_{0eR}$ and  $H_{0eR}$ are the effective couplings which originate with the gauge interaction and Yukawa interaction, respectively.
The explicit form of $G_{0eR}$ and  $H_{0eR}$ and the function $F_2(x),~F_4(x)$ are defined in Appendix.

We assume the GUT relation for gaugino masses
\begin{equation}
\frac{M_1}{g_1^2}=\frac{M_2}{g_2^2}=\frac{M_3}{g_3^2}.
\label{gugino GUT}
\end{equation}
Moreover, we use the observed CKM matrix as the form of unitary matrix for $e_{R}$ which diagonalizes $Y^{e}$
\footnote{Note that the effect of ${\cal O}(1)$ coefficients can be large as discussed in the previous section.
We also neglect a suppression factor ($\sim 15\%$) induced by electromagnetic corrections\cite{QED}.}. 
Here, only for simplicity, we take the vanishing CP phase.
In the above assumption, the number of free parameters which determine the branching ratios is 5 :
$m$, $m_3$,  the higgsino mixing parameter ($\mu$),
the $SU(2)_L$ gaugino mass ($M_2$), and the Higgs VEV ratio ($\tan\beta$).
In the following, instead of $m$ and $m_3$, we use the first two generation right-handed slepton masses 
defined as $m_{\tilde{e}_{R_{1,2}}}^2\equiv m^2+m^2_Z \cos^2 2\beta\sin^2\theta_W$ and 
the right-handed stau mass defined as $m_{\tilde{e}_{R_3}}\equiv m_3^2+m^2_Z \cos^2 2\beta\sin^2\theta_W$, which
give the eigenvalues of the slepton mass matrix in the limit in which the lepton Yukawa couplings are vanishing.

In Fig.\,1, we show the dependence of $Br(\tau \to \mu \gamma), Br(\mu \to e \gamma)$ on the first two generation 
right-handed slepton mass.
Here, we take $M_2=120\ {\rm GeV},~\mu=250\ {\rm GeV},~\tanB$ $=10,~m_{\tilde{e}_{R_3}}=150\ {\rm GeV}$
\footnote{
This parameter set has been adopted to avoid the weak scale instability in the paper\cite{KMMSSY} in which we 
discussed the little hierarchy problem.}.
The sign of $\mu$ is taken so that the chargino contribution to the $b\rightarrow s\gamma$ process has the opposite sign
for the SM contribution. 
\begin{figure}[!h]
\begin{center}
\includegraphics[width=9.6cm,clip]{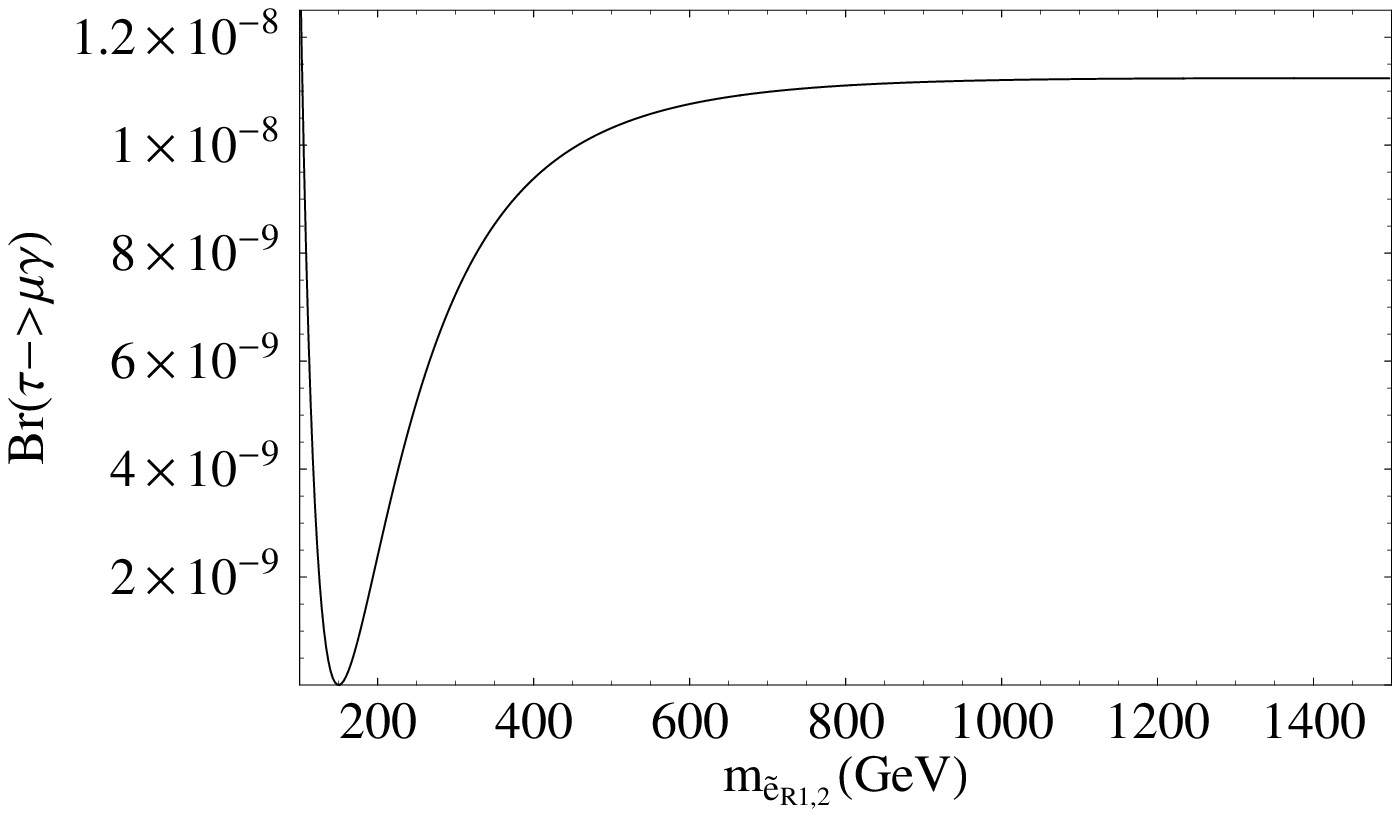}
\end{center}
\vspace{-3mm}
\begin{center}
\includegraphics[width=9.7cm,clip]{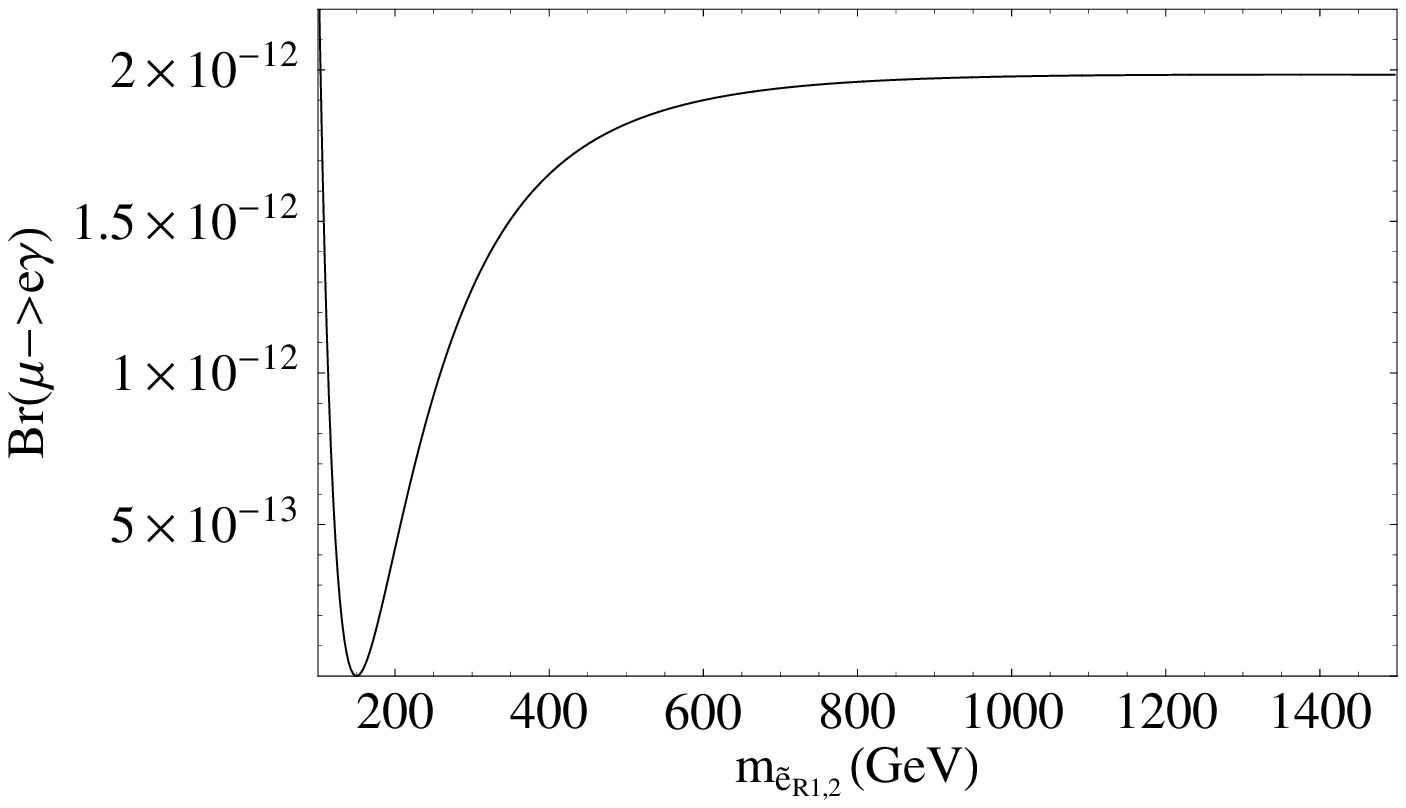}
\caption{Dependence of $Br(\tmg)$ and $Br(\meg)$ on $m_{\tilde{e}_{R_{1,2}}}$}
\label{mRdep}
\end{center}
\end{figure}
We can see that the branching ratios become zero at the point where $m_{\tilde{e}_{R_{1,2}}}$ is 150 GeV.
This is because the flavor violating source ($m^2_{mix}$) is vanished, since right-handed slepton masses are universal.
Since we do not take account of the slepton LR mixing or the radiatively induced flavor violating source in this paper,
it seems that their contributions dominate actually at this region.
On the other hand, as we can see in Fig.\,1, both branching ratios are not vanishing in the limit 
$m_{\tilde{e}_{R_{1,2}}}\gg m_{\tilde e_{R3}}$. Let us explain the reason, using the mass insertion method.
Both $\tau \to \mu \gamma$ and $\mu \to e \gamma$ diagrams (a) and (b) include the first or the second generation 
right-handed slepton in the loop. 
However, in the limit $m_{\tilde{e}_{R_{1,2}}}\gg m_{\tilde e_{R3}}$, the internal lines can be estimated as
\\
\hspace{-5mm}
\begin{tabular}{cr}
\begin{minipage}{0.5\hsize}
\begin{flushleft}
\includegraphics[width=5cm,clip]{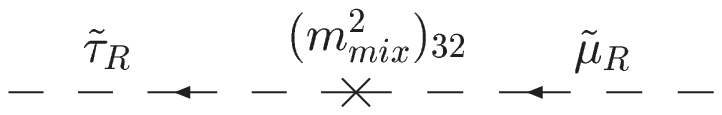}
\end{flushleft}
\end{minipage}
\begin{minipage}{0.5\hsize}
\vspace{5mm}
\begin{equation}
\hspace*{-5.0cm} \sim~\frac{1}{m^2_{\tilde{e}_{R_{3}}}} (m^2_{3}-m^2) \lambda^2 
\frac{1}{m^2_{\tilde{e}_{R_{1,2}}}}~~\xrightarrow{m~\rightarrow~\infty}~~\frac{\lambda^2}{m^2_{\tilde{e}_{R_{3}}}}~
\label{mix21}
\end{equation}
\end{minipage}
\end{tabular}
for $\tmg$, \\
\hspace{-5mm}
\begin{tabular}{cr}
\begin{minipage}{0.5\hsize}
\begin{flushleft}
\includegraphics[width=5cm,clip]{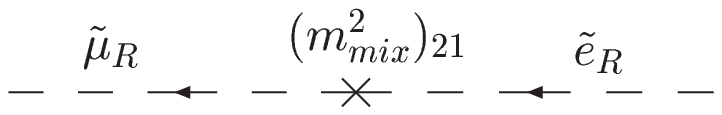}
\end{flushleft}
\end{minipage}
\begin{minipage}{0.5\hsize}
\vspace{5mm}
\begin{equation}
\hspace*{-6.0cm} \sim~\frac{1}{m^2_{\tilde{e}_{R_{1,2}}}} (m^2_{3}-m^2) \lambda^5 
\frac{1}{m^2_{\tilde{e}_{R_{1,2}}}}~~\xrightarrow{m~\rightarrow~\infty}~~0~,
\end{equation}
\end{minipage}
\end{tabular}
\vspace{-5mm}
\hspace{-5mm}
\begin{tabular}{cr}
\begin{minipage}{0.5\hsize}
\begin{flushleft}
\includegraphics[width=6.5cm,clip]{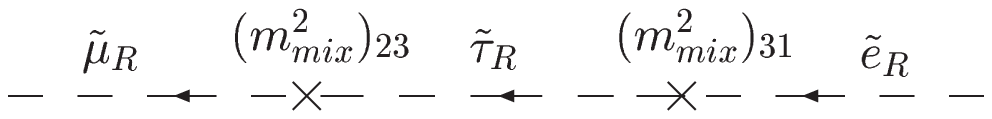}
\end{flushleft}
\end{minipage}
\begin{minipage}{0.5\hsize}
\begin{equation}
\hspace*{-1.8cm} \sim~(m^2_{3}-m^2)^2 \lambda^5 \frac{1}{m^4_{\tilde{e}_{R_{1,2}}}} 
\frac{1}{m^2_{\tilde{e}_{R_{3}}}} 
~\xrightarrow{m~\rightarrow~\infty}~\frac{\lambda^5}{m^2_{\tilde{e}_{R_{3}}}}~
\label{mix231}
\end{equation}
\end{minipage}
\end{tabular}
\\
\vspace{8mm}
\\
for $\meg$.
When $m_{\tilde{e}_{R_{1,2}}}\gg m_{\tilde e_{R3}}$,
the flavor changing off-diagonal entries become also large.
As a result, both transition rates remain finite. From Fig.\,1,
the branching ratios no longer depend on the first two generation masses 
in $m_{\tilde{e}_{R_{1,2}}}>600$ GeV region.

In Figs.\,2 and \,4, we show the $\mu$ dependence of $Br(\tau \to \mu \gamma)$ and $Br(\mu \to e \gamma)$.
Here, we take $m_{\tilde{e}_{R_{1,2}}}=1000$ GeV and the other parameters are taken as in the previous one.
We find that around $\mu \sim 20, -1000$ GeV the branching ratios become very small.
To understand this behavior, we should remind that neutralino amplitude consists of two parts.
One is proportional to $m_{l_i}$, the other is proportional to $M_{\tilde{\chi}^0_a}$,
they correspond to diagram (a) and diagram (b), respectively.
In Figs.\,3 and \,5, the dotted line represents the amplitude which are proportional to $m_{l_j}$ and the 
solid line represents the amplitude which is proportional to $M_{\tilde{\chi}^0_a}$.
As we can see from these figures, the $m_{l_j}$ part depends hardly on $\mu$. 
This is because this part comes mainly from the diagram (a) in which higssinos do not contribute.
On the other hand, the $M_{\tilde{\chi}^0_a}$ part which comes mainly from the diagram (b)  depends on $\mu$.
This part changes its sign from $\mu<0$ to $\mu>0$ and is decreasing as increasing $\mu$ due to the decoupling.
Basically, contribution of diagram (b) is larger than that of diagram (a) due to
$\tan\beta$ enhancement.
However, when diagram (b) changes its sign, or it goes to decouple as increasing $|\mu|$,
there are two points where the magnitude of diagram (b) becomes smaller than that of diagram (a).
If each diagrams have different sign in these points, they are cancelled.
Now, we could understand the vanishing point in Figs.\,2 and \,4.
Note that the cancellation happens mainly in the $\mu<0$ region.
Consequently, the branching ratios do not strongly depend on $\mu$ in the positive $\mu$ region
where the chargino contribution to $b \to s \gamma$ has opposite sign for the SM contribution and the charged Higgs
contribution to $b \to s \gamma$.
\begin{figure}[!h]
\begin{center}
\includegraphics[width=6cm,clip]{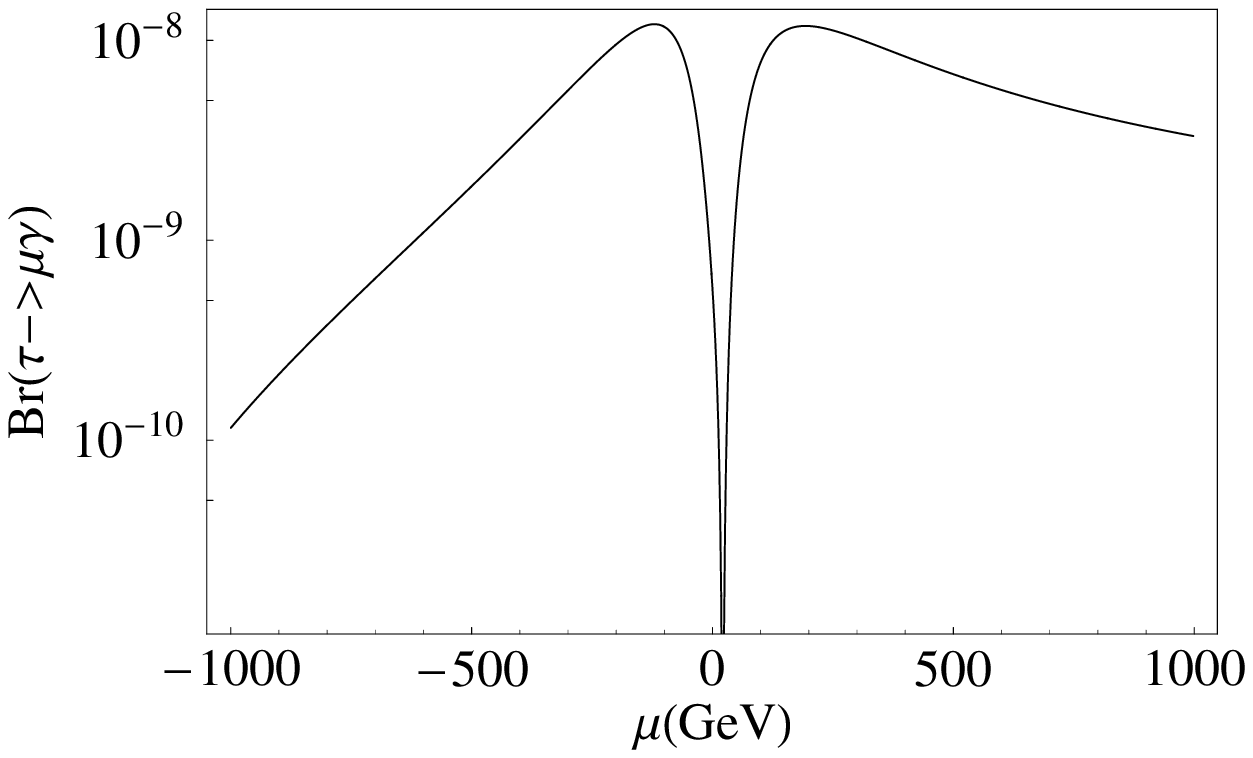}
\caption{\small $\mu$ dependence of $Br(\tmg)$}
\label{Bt-mu}
\end{center}
\begin{center}
\includegraphics[width=6cm,clip]{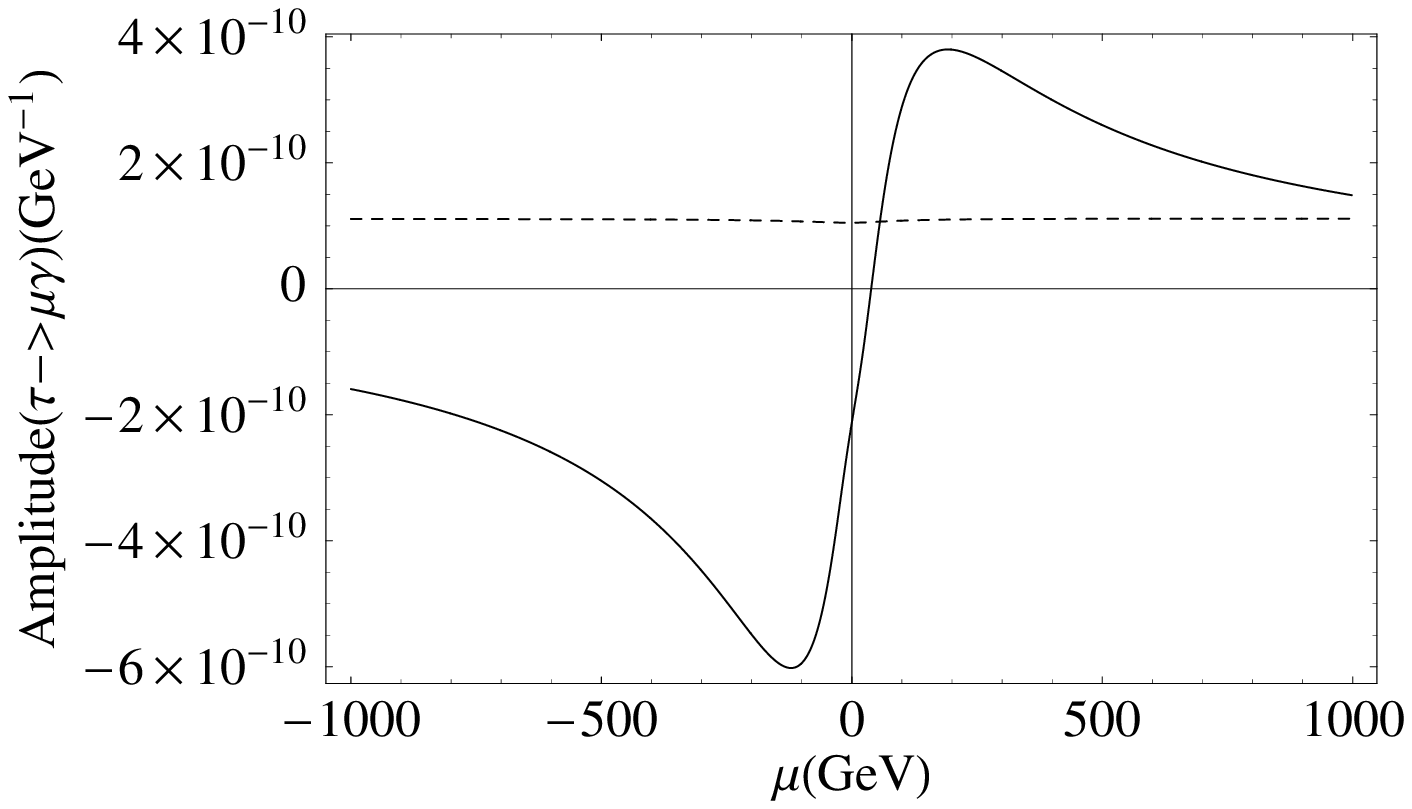}
\caption{\small The solid line represents the $\mu$ dependence of the $\tmg$ neutralino amplitude which is proportional to $m_{\xn}$, and the 
dotted line represents the $\mu$ dependence of the $\tmg$ neutralino amplitude which is proportional to $m_{\tau}$.}
\label{At-mu}
\end{center}
\end{figure}
\begin{figure}[!h]
\begin{center}
\includegraphics[width=6.4cm,clip]{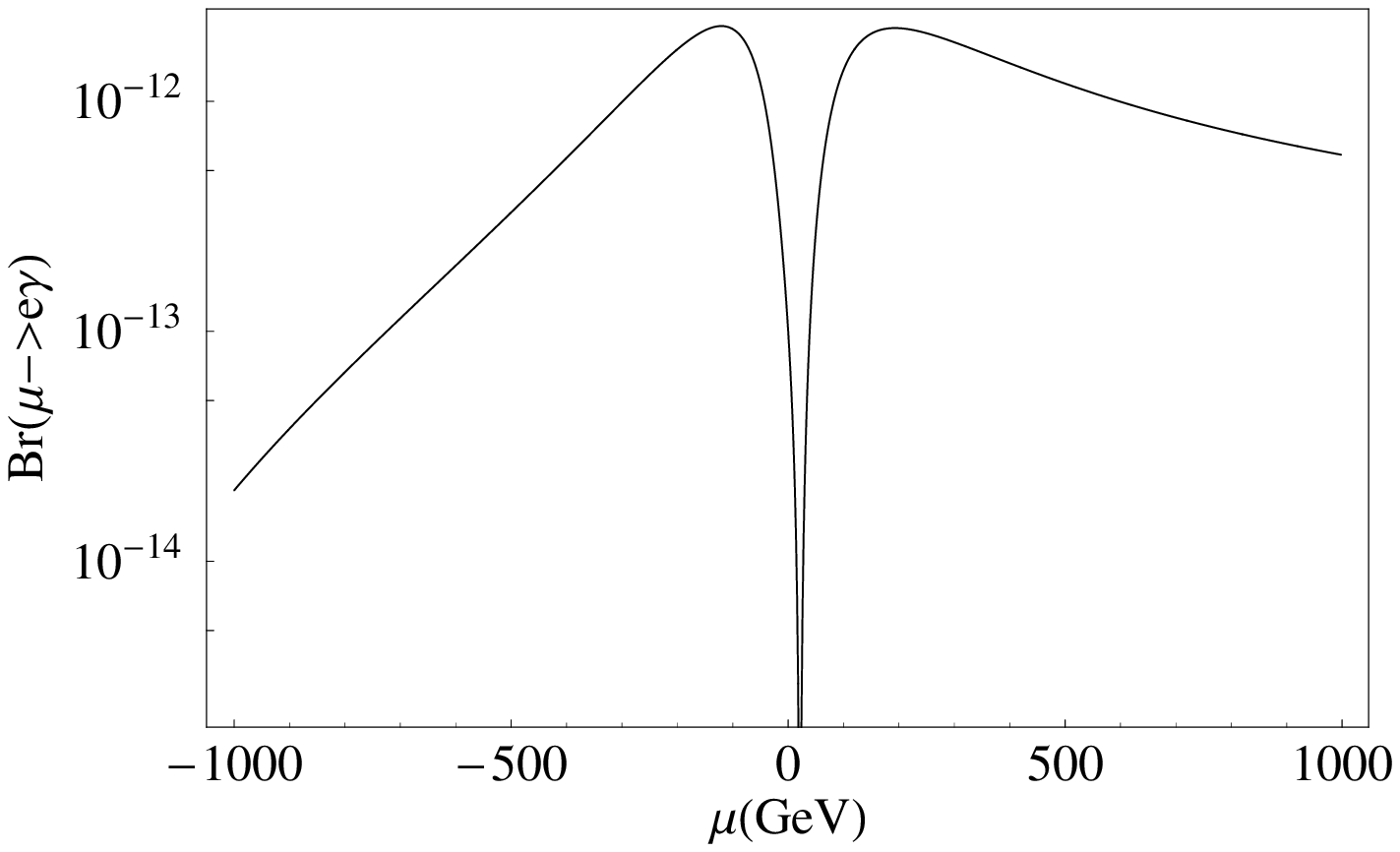}
\caption{\small The $\mu$ dependence of $Br(\meg)$}
\label{Bm-mu}
\end{center}
\begin{center}
\includegraphics[width=6.4cm,clip]{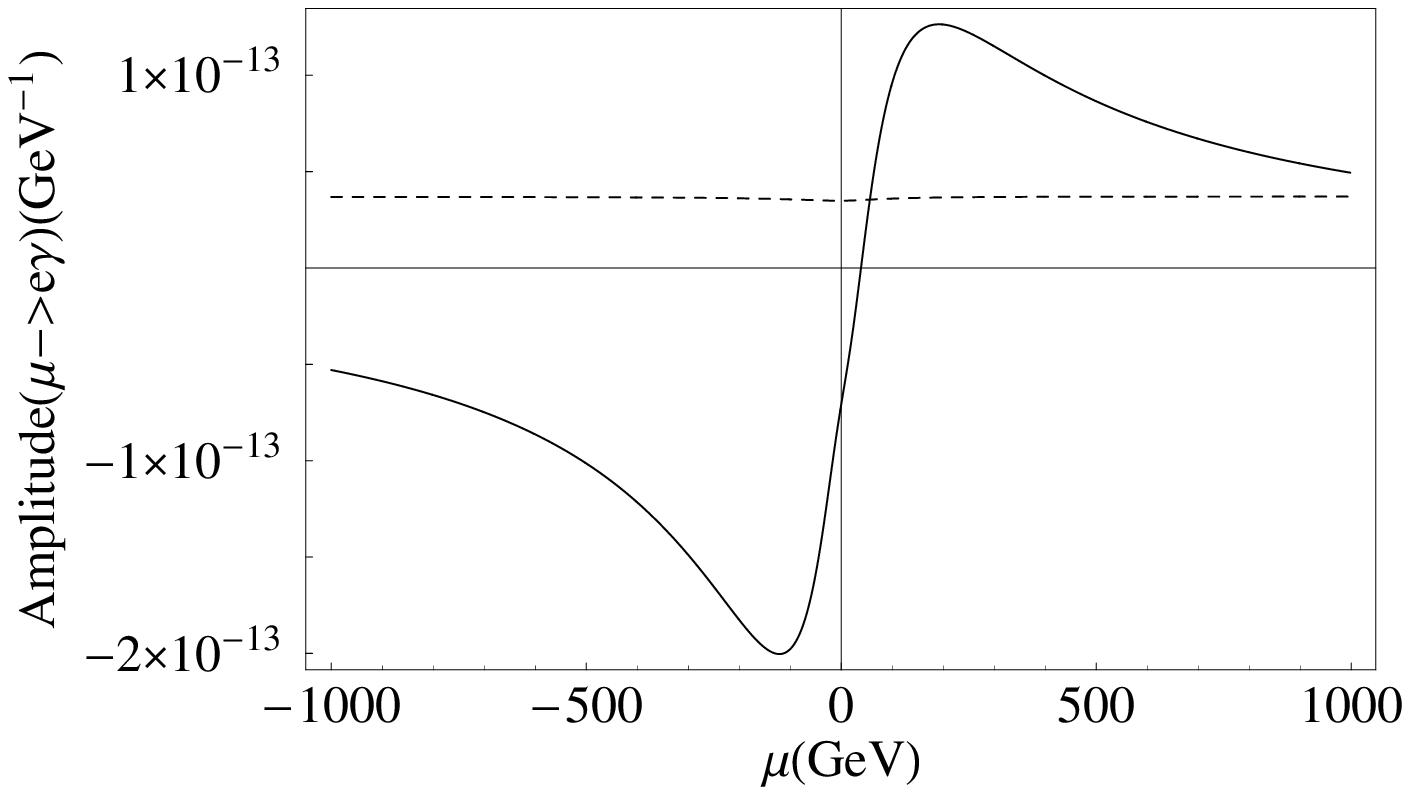}
\caption{\small The solid line represents the $\mu$ dependence of the $\meg$ neutralino amplitude which is proportional to $m_{\xn}$, and the 
dotted line represents the $\mu$ dependence of the $\meg$ neutralino amplitude which is proportional to $m_{\mu}$.}
\label{Am-mu}
\end{center}
\end{figure}

In Figs.\,6, we show the $M_2$ dependence of the branching ratios when $\mu=250$ GeV.
Here, we take the same parameters as those in the previous figures.
We can see the point around $M_2 \sim 0$ GeV in which the branching ratios are vanishing.
This is because of the cancellation between the contributions from the diagram (a) and (b) as in Figs.\,2 and \,4.
The region with $M_2<0$ and $\mu>0$  corresponds to the region $M_2>0$ and $\mu<0$. 
\begin{figure}[h!]
\begin{center}
\includegraphics[width=6.4cm,clip]{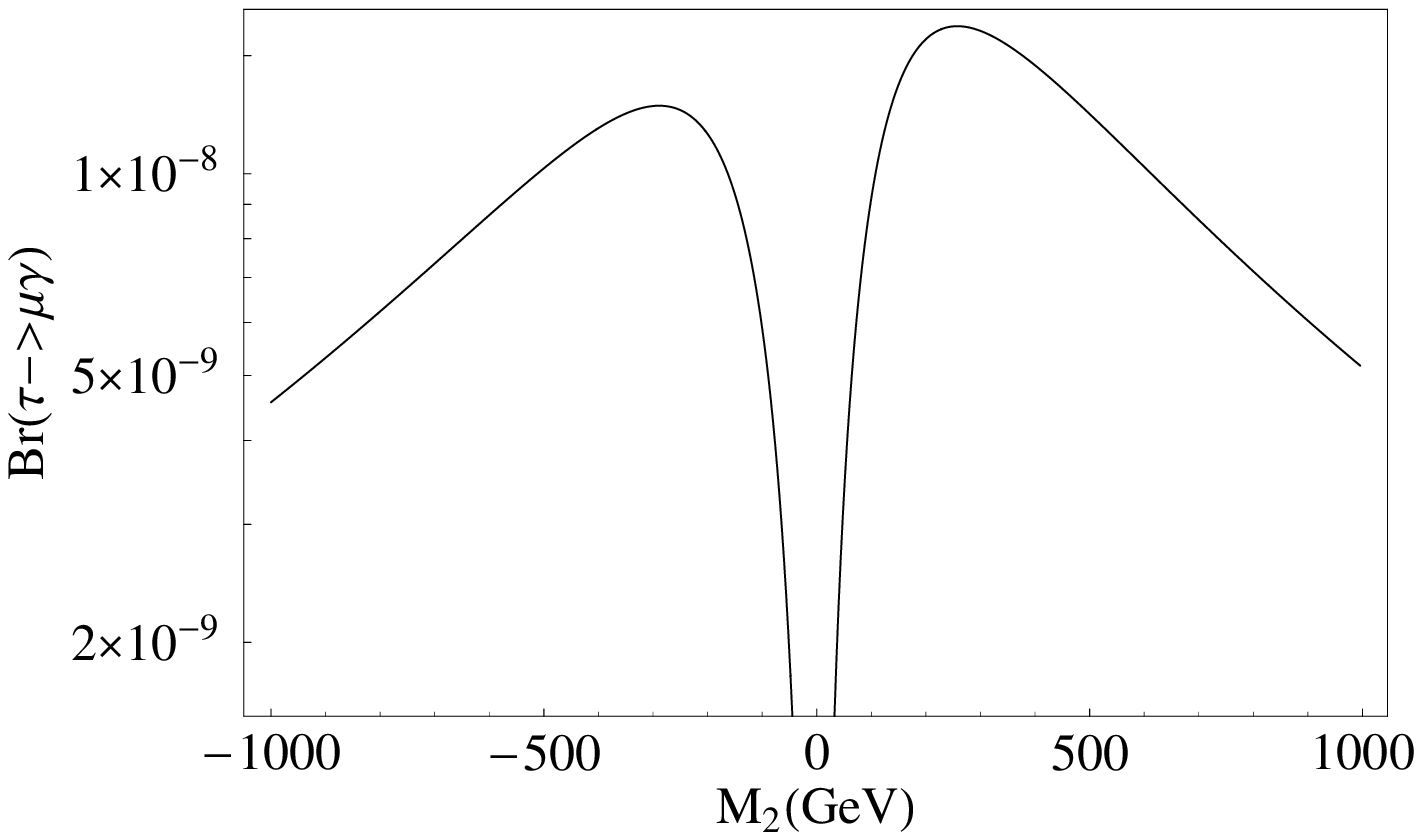}
\end{center}
\vspace{-3mm}
\begin{center}
\includegraphics[width=6.4cm,clip]{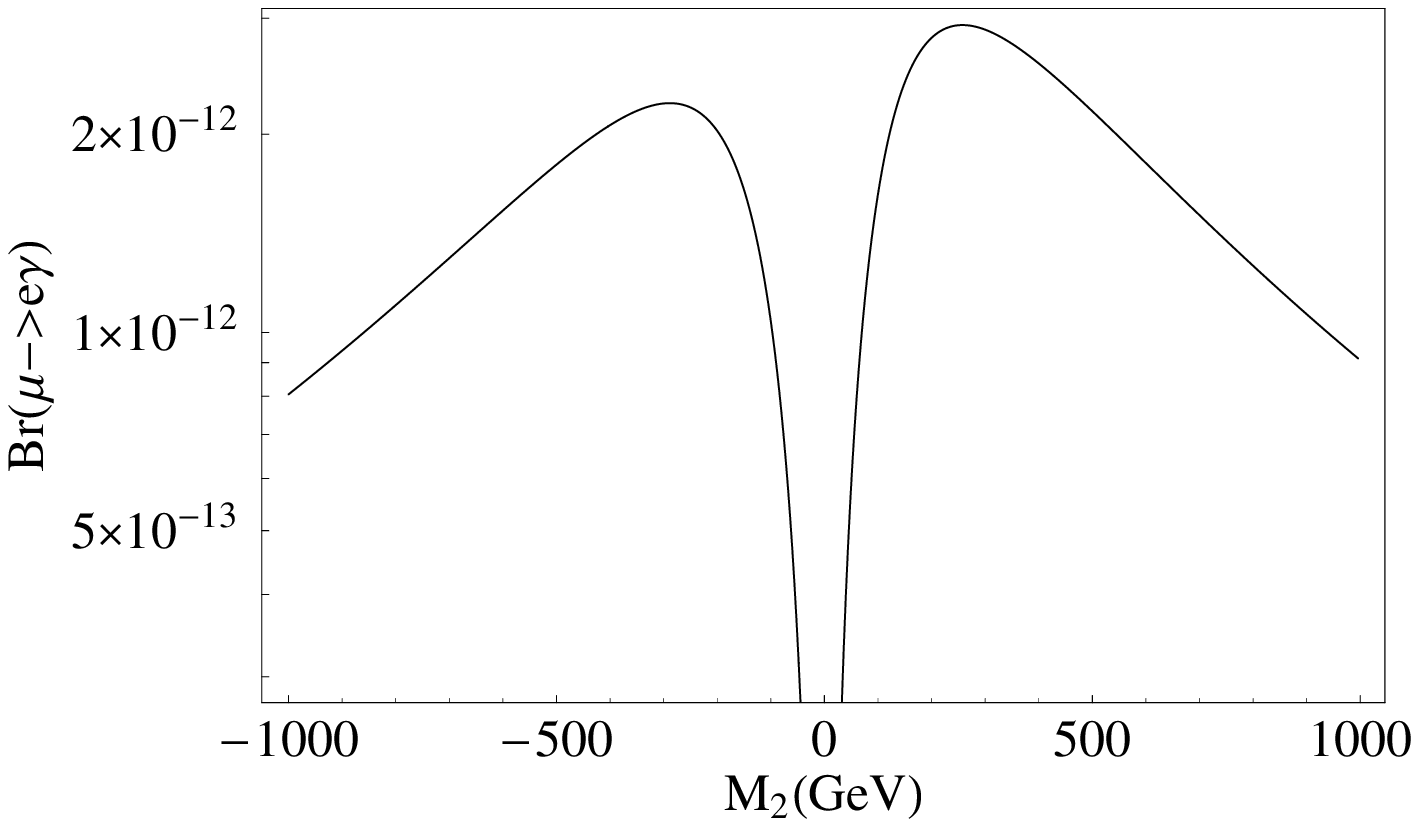}
\caption{\small $M_2$ dependence of $Br(\tmg)$ and $Br(\meg)$}
\label{Bt-M2}
\end{center}
\end{figure}

Figs.\,7 and \,8 show the $\tau \to \mu \gamma$ and $\mu \to e \gamma$ branching
ratios for $\tan\beta$ and right-handed stau mass when $M_2=120$ GeV. 
We take the other parameters to be the same as in the previous figures.
In Fig.\,7, the dashed line represents $Br(\tau \to \mu\gamma)=4.5 \times 10^{-8}$ and
the left side of the dashed line is excluded by current experiment. The solid lines represents
$Br(\tau \to \mu\gamma) = 1 \times 10^{-8}, 10^{-9}, 10^{-10}, 10^{-11}$ from the up-left side.
It is found that if right-handed stau mass is about 100 GeV, $\tan\beta$ cannot be taken larger than 9.       
The region where $4.5\times 10^{-8}>Br(\tau \to \mu \gamma) > 5\times 10^{-9}$ is the target region of the B-factory 
or super B-factory\cite{SuperB}.
If the right-handed stau is light ($m_{\tilde e_{R3}}<200$ GeV, when $\tanB\sim 10$), 
the $\tau \to \mu \gamma$ process can be 
discovered at the B factory or super B-factory.
In Fig.\,8, the solid lines represent $Br(\mu \to e \gamma)= 10^{-11},\,10^{-12},\,10^{-13}, \,10^{-14}, 10^{-15}$ 
from the up-left side 
and the dashed
line represents the current limit $Br(\mu\to e\gamma)<1.2\times 10^{-11}$. 
The region $10^{-11}>Br(\mu \to e \gamma)>10^{-14}$ is the target region of the MEG experiment\cite{MEGe}.
Thus, we find if the right-handed stau is light ($m_{\tilde e_{R3}}<420$ GeV, when $\tanB\sim 10$), 
$\mu \to e \gamma$ may be discovered at the MEG experiment.
We calculated these figures with the negative $\mu=-250$ GeV. The results are similar to the Figs.\,7 and \,8.
However, when the $\mu$ is negative, the chargino contribution to the $b\to s\gamma$ process  
has the same sign as the SM contribution and the charged Higgs contribution. Therefore, other
large contribution to the $b\to s\gamma$ process with opposite sign is required. 

Comparing these figures, we find that the $\tau\to\mu\gamma$ process gives more severe constraints to this scenario than the 
$\mu\to e\gamma$ process at present. Before the MEG experiment finds the signal of the $\mu\to e\gamma$, 
the $\tau\to\mu\gamma$ events can be found in the B-factory.

\begin{figure}[h!]
\vspace{5mm}
\begin{center}
\includegraphics[width=8.4cm,clip]{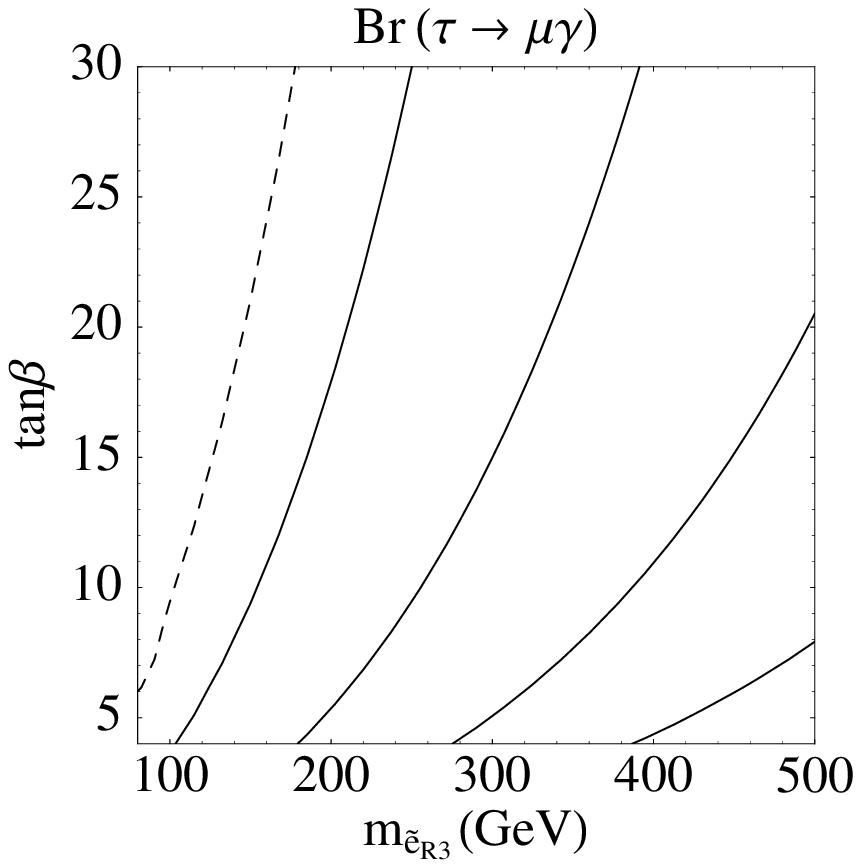}
\caption{\small Dependence of $Br(\tmg)$ on $\tanB$ and $m_{\tilde e_{R3}} (\mu=250$ GeV$)$. The solid lines represent 
$\Brt=10^{-8},~10^{-9},~10^{-10}, 10^{-11}$ from the up-left side to the down-right side. 
The dashed line represents the current bound.}
\label{Brt.m3.tanB}
\end{center}
\vspace{7.3mm}
\begin{center}
\includegraphics[width=8.4cm,clip]{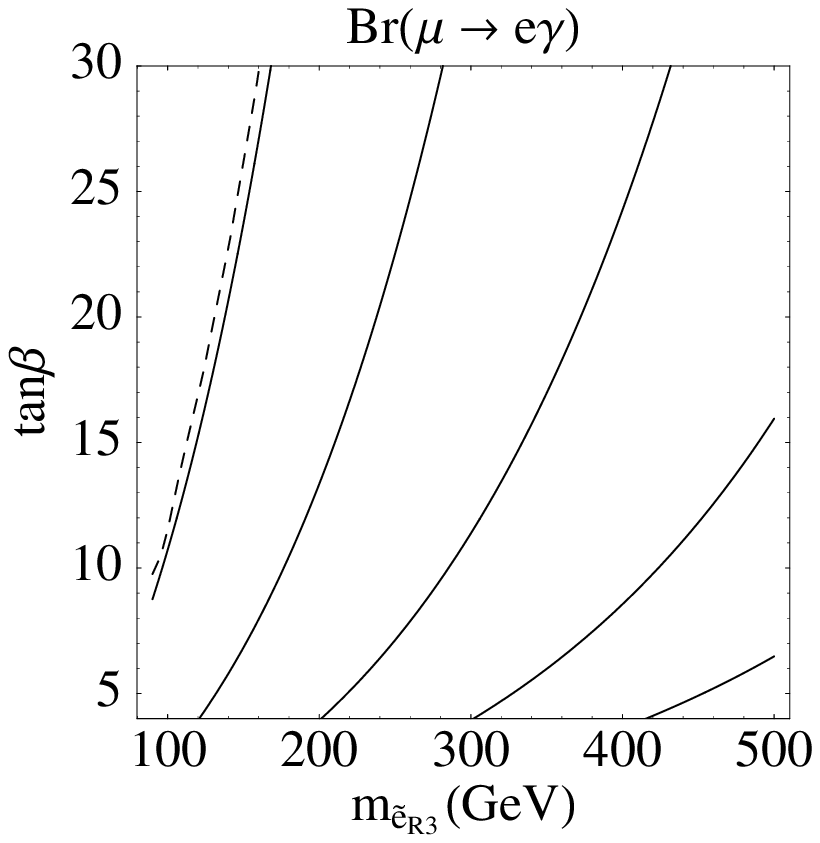}
\caption{\small Dependence of $Br(\meg)$ on $\tanB$ and $m_{\tilde e_{R3}} (\mu=250$ GeV$)$. The solid lines represent 
$\Brm=10^{-11}, 10^{-12}, 10^{-13}, 10^{-14}, 10^{-15}$ from the up-left side to the down-right. 
The dashed line represents the current bound.}
\label{Brm.m3.tanB}
\end{center}
\end{figure}

\section{Conclusion}
In this paper, we have analyzed the lepton flavor violating $\tau \to \mu \gamma$ and $\mu \to e \gamma$ processes
in SUSY GUT model in which the third generation sfermion masses, $m_{3}$ of ${\bf 10}$ of SU(5) have different masses
from the others, $m$. Such mass spectrum is consistent with the present constraints from the various FCNC processes.
And it is interesting that such mass spectrum is naturally derived from the GUT models with 
${\rm E}_6 \times $ Horizontal Symmetry.

This mass spectrum allowed us to take very large value for $m$ without destabilizing the weak scale in the MSSM,
because stop masses are determined by $m_{3}$.
We have considered a situation where $m_3$ is of order the weak scale 
while $m$ is larger than the weak scale.
In this case, the LFV decay rates are determined mainly by the right-handed stau mass and $\tan\beta$.
Especially, the branching ratios of $\tau \to \mu \gamma$ and $\mu \to e \gamma$ strongly depend on the 
right-handed stau mass. 
This means that this type of soft mass matrix can be tested by checking the relation among the LFV decay rates
and the right-handed stau mass which may be measured in the direct search, though the ambiguity due to the ${\cal O}(1)$
coefficients still exists. 
As we can see from Figs.\,7 and \,8, if the right-handed stau is lighter than 420 GeV (when $\tanB\sim 10$), 
$\mu \to e \gamma$ process
may be discovered at the MEG experiment. It may be interesting that the B-factory can find the $\tau\to\mu\gamma$ events before the MEG experiment
find the $\mu\to e\gamma$ events because the present constraints from $\tau\to\mu\gamma$ is more severe than that from the 
$\mu\to e\gamma$ process. 

Moreover, this model provides some interesting predictions.
First, this model says that the final state lepton tends to have the right-handed chirality, 
because the intermediate state must have
the right-handed chirality to pick up the flavor violating mass matrix entries.
On the other hand, the LFV processes induced radiatively via large Dirac neutrino Yukawa interactions predicts
an opposite chirality (left-handed) for the final state lepton. 
This difference can be checked by measuring the angular distribution of the final lepton for the initial lepton's spin.
Second, this model predicts several relations among $\tau \to \mu \gamma$, $\mu \to e \gamma$, and $\tau\to e\gamma$ as
$Br(\tau \to \mu \gamma) \sim 10^{3} \times [{\cal O}(1)]^4 Br(\mu\to e\gamma)\sim 20\times [{\cal O}(1)]^4 Br(\tau \to e \gamma)$.
This enhancement factor of the former relation can be understood as $Br(\tau \to e \nu \bar{\nu}) \times \lambda^4/\lambda^{10}$ and
that of the latter as $\lambda^{-2}$.
It is important to test these predictions in future experiments.

\section*{Acknowledgements}
N.M. and K.S. are supported in part by Grants-in-Aid for Scientific 
Research from the Ministry of Education, Culture, Sports, Science 
and Technology of Japan. The work of T.Y. is supported by the 21st Century COE
Program of Nagoya University.

\section*{Appendix}

In this appendix, we give our notation and conventions adopted in Section 3, which are basically the same as in the papers
\cite{rules1,rules2}.

First we consider the sfermions.
Below the weak scale, the mass matrix of the charged slepton
is given in the super-CKM basis as
\begin{equation}
-{\cal L}_m \ni
\begin{pmatrix} \tilde{e}^{\dagger}_L & \tilde{e}^{\dagger}_R\end{pmatrix}
m^2_{\tilde{e}}
\begin{pmatrix} \tilde{e}_L \\ \tilde{e}_R \end{pmatrix}\,,
\end{equation}
\begin{equation}
m^2_{\tilde{e}} =
\begin{pmatrix}
m^2_{\bf\bar 5}+m^2_Z \cos^2 2\beta(-\frac{1}{2}+\sin^2\theta_W)+m_e^2 & -m_e (A_e + \mu \tan\beta) \cr
-m_e (A_e + \mu \tan\beta) & V_{10}^{\dagger} m_{\bf 10}^2 V_{10} + m^2_Z \cos^2 2\beta\sin^2\theta_W+m_e^2 \cr
\end{pmatrix}\,,
\end{equation} 
where $m_{\bf\bar 5}^2$ and $m_{\bf 10}^2$ are slepton soft mass matrices given in 
Eq.~(\ref{special mass}) at the low energy, 
$V_{10}$ is a diagonalizing matrix given in Eq.~(\ref{V10}), and $m_e$ is the diagonal charged lepton mass matrix. 
Here we assumed that the $A$-term is proportional to the Yukawa matrix.
We define the $6\times 6$ unitary matrix $\Gamma_{e}$ which rotates the left(right)-handed slepton fields into
mass eigenstate $\tilde{e}_k~(k=1,\ldots,6)$ as 
\begin{equation}
\Gamma_e (m^2_{\tilde{e}}) \Gamma_e^{\dagger} =({\rm diagonal}),~~~~~~
\tilde{e}_{L(R)}  = \Gamma^{\dagger}_{eL(R)} \tilde{e}, \nonumber \\
\end{equation}
where $6\times 3$ matrices $\Gamma_{eL(R)}$ are defined as
\begin{equation}
\Gamma_{eL}^{ki}=\Gamma_{e}^{ki},~~~~\Gamma_{eR}^{ki}=\Gamma_{e}^{k,i+3}\,.
\end{equation}

Now we turn to the neutralinos.
The mass term of the neutralino sector is given by
\begin{equation}
-{\cal L}_m \ni \frac{1}{2}
\begin{pmatrix} \tilde{\gamma}& \tilde{Z}& \tilde{H_d^0}& \tilde{H_u^0} \end{pmatrix}
\cdot {\bf M}_{\xn}
\begin{pmatrix} \tilde{\gamma}\\ \tilde{Z}\\ \tilde{H_d^0}\\ \tilde{H_u^0} \end{pmatrix}
+h.c.\,,
\end{equation}
where
\begin{equation}
{\bf M}_{\xn} =
\begin{pmatrix}
M_1 \cos^2\theta_W + M_2 \sin^2\theta_W& \cos{\theta_W}\sin{\theta_W}(M_2-M_1)& 0& 0 \cr
\cos{\theta_W}\sin{\theta_W}(M_2-M_1)& M_1\sin^2\theta_W + M_2\cos^2\theta_W& m_Z \cos\beta & -m_Z \sin\beta \cr
0& m_Z\cos\beta& 0& -\mu \cr
0& -m_Z\sin\beta& -\mu& 0
\end{pmatrix}\,.
\label{photinomat}
\end{equation}
This neutralino mass matrix can be diagonalized by a real orthogonal matrix N:
\begin{equation}
{\bf N} {\bf M}_{\xn} {\bf N}^{\rm T} = ({\rm diagonal})\,. 
\end{equation}
The mass eigenstates $\xn_i$ $(i=1,...,4)$ are given by
\begin{equation}
\begin{pmatrix}
\xn_1\\ \xn_2\\ \xn_3\\ \xn_4
\end{pmatrix}
={\bf N}
\begin{pmatrix}
\tilde{\gamma}\\ \tilde{Z}\\ \tilde{H}_d^0\\ \tilde{H}_u^0
\end{pmatrix}\,.
\end{equation}

We use the following vertices.
 
\begin{tabular}{cr}
\begin{minipage}{0.5\hsize}
\begin{flushleft}
\includegraphics[width=3.5cm,clip]{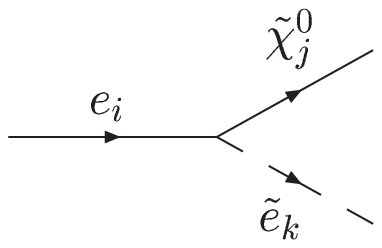}
\end{flushleft}
\end{minipage}
\begin{minipage}{0.5\hsize}
\begin{equation}
\hspace*{-8cm}=~-ig \Bigl[( \sqrt{2} G_{0eL}^{jki}+ H_{0eR}^{jki}) P_L
-(\sqrt{2} G_{0eR}^{jki}- H_{0eL}^{jki})P_R \Bigr]
\end{equation}
\end{minipage}\\[8mm]
\begin{minipage}{0.5\hsize}
\begin{flushleft}
\includegraphics[width=3.5cm,clip]{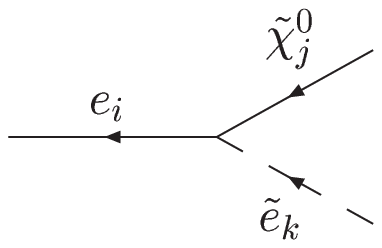}
\end{flushleft}
\end{minipage}
\begin{minipage}{0.5\hsize}
\begin{equation}
\hspace*{-7.7cm}=~-ig \Bigl[ (\sqrt{2} G_{0eL}^{*jki}+H_{0eR}^{*jki}) P_R
-(\sqrt{2} G_{0eR}^{*jki}-H_{0eL}^{*jki}) P_L \Bigr],
\end{equation}
\end{minipage}
\end{tabular}
where
\begin{eqnarray}
G_{0eL}^{jki} &\equiv&
\Bigl[-\sin\theta_W {\bf N}^{j1*} + \frac{1}{\cos\theta_W}(-\frac{1}{2}+\sin^2\theta_W)
{\bf N}^{j2*} \Bigr]\Gamma_{eL}^{ki}\,,\\ 
G_{0eR}^{jki} &\equiv&
\Bigl[-\sin\theta_W {\bf N}^{j1} + \frac{\sin^2\theta_W}{\cos\theta_W}
{\bf N}^{j2} \Bigr]\Gamma_{eR}^{ki}\,,\\ 
H_{0eL}^{jki} &\equiv& {\bf N}^{*j3} (\Gamma_{0eL} Y_e)^{ki}/g\,,\\
H_{0eR}^{jki} &\equiv& {\bf N}^{*j3} (\Gamma_{0eR} Y_e)^{ki}/g\,.
\end{eqnarray}

Finally, we show the functions $F_2$ and $F_4$ which appear in eq.\,(\ref{MassEigenAmp}) as
\begin{eqnarray}
F_2(x)&=&\frac{1}{12(x-1)^4}(2x^3 +3x^2 -6x +1 -6x^2 \log x )\,, \nonumber \\
F_4(x)&=&\frac{1}{2(x-1)^3}(x^2 -1 -2x \log x )\,.
\end{eqnarray}

\end{document}